\documentclass[12pt]{iopart}

\usepackage{amssymb}
\usepackage{amsthm}

 \newtheorem{theorem}{Theorem}[section]
       \newtheorem{lemma}[theorem]{Lemma}

    \newtheorem{definition}[theorem]{Definition}
       
\newcommand{\ud}{\mathrm{d}}

\begin{document}


\title[]{Gelfand-Yaglom-Perez Theorem for Generalized Relative Entropies}

\author{Ambedkar Dukkipati\footnote{Corresponding author}, Shalabh
Bhatnagar and M Narasimha Murty}

\address{Department of Computer Science and Automation,
Indian Institute of Science, Bangalore-560012, India.}
\ead{\mailto{ambedkar@csa.iisc.ernet.in},
\mailto{mnm@csa.iisc.ernet.in}, \mailto{shalabh@csa.iisc.ernet.in}}

\begin{abstract}
	The measure-theoretic definition of Kullback-Leibler
	relative-entropy 
	(KL-entropy) plays a basic role in the  
	definitions of
	classical information measures. Entropy, mutual information
	and conditional forms of entropy can be expressed in terms
	of KL-entropy and hence properties of their measure-theoretic
	analogs will follow from those of
	measure-theoretic KL-entropy. These measure-theoretic
	definitions are key to extending the ergodic
	theorems of information theory to non-discrete cases. A
	fundamental theorem in this respect is the  
	Gelfand-Yaglom-Perez (GYP) Theorem (Pinsker, 1960,
	Theorem. 2.4.2) which states that measure-theoretic relative-entropy equals 
	the supremum of relative-entropies over all measurable
	partitions. This paper states and proves
	the GYP-theorem for R\'{e}nyi relative-entropy of order
	greater than one. Consequently, the result can be easily
	extended to Tsallis 
	relative-entropy. 
\end{abstract}

\maketitle

\section{Introduction}
\label{Section:Introduction}
	\noindent
	R{\'{e}}nyi~\cite{Renyi:1960:SomeFundamentalQuestionsOfInformationTheory},
	by replacing linear     
        averaging in Shannon entropy with Kolmogorov-Nagumo
        average or quasilinear mean and further
        imposing the additivity   
        constraint, proposed a one-parameter family 
        of measures of information ($\alpha$-entropies) which is
        defined as follows:
        \begin{equation}
	\label{Equation:Definition_RenyiEntropy}
        S_{\alpha}(p) = \frac{1}{1-\alpha} \ln \left(\sum_{k=1}^{n}
        p_{k}^{\alpha} \right) \enspace,
        \end{equation}
	where $p=\{p_{k}\}_{k=1}^{n}$ is a probability mass function
        (pmf) and $\alpha \in \mathbb{R}$ and $\alpha >0$.
	 R\'{e}nyi entropy (\ref{Equation:Definition_RenyiEntropy}) is
        a 
        one-parameter generalization of Shannon entropy in the sense
        that the limit $\alpha \rightarrow 1$ in
        (\ref{Equation:Definition_RenyiEntropy}) retrieves Shannon
        entropy. $S_{\alpha}$ is referred as the entropy of order $\alpha$.
        Despite its formal origin, R\'{e}nyi entropy proved important
        in a variety of practical applications in coding
        theory~\cite{AczelDaroczy:1975:OnMeasuresOfInformationAndTheirCharacterization},
        statistical
        inference~\cite{ArimitsuArimitsu:2001:AnalysisOfTurbulence_SecondaryRef},
        quantum  
        mechanics~\cite{MaassenUffink:1988:GeneralizedEntropicUncertaintyRelations},
        and chaotic dynamical
        systems~\cite{HalseyJensenKadanoffProcacciaShraiman:1986:FractalMeasuresAndThierSingularities}.

	Along similar lines, R\'{e}nyi defined a one parameter
	generalization of Kullback-Leibler relative-entropy
	as~\cite{Renyi:1960:SomeFundamentalQuestionsOfInformationTheory} 
	\begin{equation}
	S_{\alpha}(p \| r) = \frac{1}{\alpha -1} \ln \sum_{k=1}^{n}
	\frac{p_{k}^{\alpha}}{r_{k}^{\alpha -1}} \enspace
 	\end{equation}
	for pmfs $p$ and $r$.

        On the other hand, though Shannon measure of entropy or
        information was developed 
        essentially for the 
        case when the random variable takes a finite number of
        values, 
        in the literature, one often encounters an extension
        of Shannon entropy in the discrete
        case to the case
        of a 
        one-dimensional random variable with density function $p$
        in the form (e.g
        ~\cite{ShannonWeawer:1949:TheMathematicalTheoryOfCommunication,Ash:1965:InformationTheory})  
        \begin{equation}
        \label{Equation:DifferentialEntropy}
          S(p) = - \int_{- \infty}^{+ \infty} p(x) \ln p(x)\, \ud x \enspace.
        \end{equation}
	(\ref{Equation:DifferentialEntropy}) is known as differential
        entropy in information theory and Boltzmann H-function in Physics.
        Indeed, during the early stages of development of
        information theory, the important paper
        by Gelfand, Kolmogorov and
        Yaglom~\cite{GelfandKolmogorovYaglom:1956:OnTheGeneralDefinitionOfTheAmountOfInformation}   
        called attention to the case where entropy is defined on an
        arbitrary measure space $(X, \mathfrak{M},\mu)$. In this
        respect, Shannon entropy of a 
        probability density function $p:X 
        \rightarrow {\mathbb{R}}^{+}$ can be defined as
        \begin{equation}
         \label{Equation:ME:ShannonEntropyOf-pdf} 
        S(p) = - \int_{X} p \ln p \, \ud \mu \enspace,
        \end{equation}
        provided the integral on right exists.
        One can see from the above definition that the concept of
        ``entropy of a pdf'' is a misnomer: there
        is always another measure $\mu$ in  the background. In the
        discrete case considered by Shannon, $\mu$ is the cardinality
        measure\footnote{Counting or cardinality measure $\mu$ on a
          measurable space $(X,\mathfrak{M})$, when is $X$ is a
          finite set and $\mathfrak{M} = 2^{X}$, is defined as $\mu(E)
          = \# E$, $\forall E \in \mathfrak{M}$.}~\cite[pp.19]{ShannonWeawer:1949:TheMathematicalTheoryOfCommunication};
        in the continuous case considered by both Shannon and Wiener,
        $\mu$ is the Lebesgue
        measure
        cf.~\cite[pp.54]{ShannonWeawer:1949:TheMathematicalTheoryOfCommunication} 
        and 
        \cite[pp.61, 62]{Wiener:1948:Cybernetics}. All entropies are
        defined with respect to some measure
        $\mu$,
        as Shannon and Wiener both emphasized in~\cite[pp.57,
        58]{ShannonWeawer:1949:TheMathematicalTheoryOfCommunication} 
        and~\cite[pp.61, 62]{Wiener:1948:Cybernetics} respectively.

        This case was studied independently
        by
        Kallianpur~\cite{Kallianpur:1960:OnTheAmountOfInformationContainedInASingmaField} 
        and Pinsker~\cite{Pinsker:1960:InformationAndInformationStability},
        and perhaps others were guided by the earlier work
        of Kullback and Leibler~\cite{KullbackLeibler:1951:OnInformationAndSufficiency},
        where one would define entropy in terms of Kullback-Leibler
        relative-entropy.

	In this respect
        Gelfand-Yaglom-Perez theorem
        (GYP-theorem)~\cite{GelfandYaglom:1959:CalculationOfTheAmountOfInformation_Etc,Perez:1959:InformationTheoryWithAbstractAlphabets,Dobrushin:1959:GeneralFormulationsOfShannonsbasicTheorems}
        plays an important role, which equips measure-theoretic
        KL-entropy with a fundamental definition.
	The main contribution of this paper is to state and prove GYP-theorem for
	R\'{e}nyi relative entropy of order $\alpha >1$.

        We review the measure-theoretic formalisms for classical
        information measures in
        \S~\ref{Section:ClassicalInformationMeasures},
        where we discuss the relation between Shannon entropy and
        KL-entropy in the measure-theoretic case.
        We extend measure-theoretic definitions to generalized
        information measures in
        \S~\ref{Section:MeasureTheoreticDefinitionsOfGeneralizedInformationMeasures}.  
	Finally, Gelfand-Yaglom-Perez
        theorem in the general case is presented in
        \S~\ref{Section:GelfandYaglomPerez_Theorem}.


\section{Measure Theoretic Definitions of Classical Information Measures}
\label{Section:ClassicalInformationMeasures}
	\noindent
        Let $(X,\mathfrak{M},\mu)$ be a measure space. $\mu$
        need not be a probability measure unless otherwise specified.
        Symbols $P$, $R$ will denote probability measures on
        measurable space $(X,\mathfrak{M})$ and $p$, $r$  
        denote $\mathfrak{M}$-measurable functions on $X$.
        An $\mathfrak{M}$-measurable function $p:X \rightarrow
        {\mathbb{R}}^{+}$ is said to be a probability 
        density function (pdf) if $\int_{X} p \, \ud \mu = 1$.

        In this general setting, entropy $S(p)$ of pdf $p$
        defined in (\ref{Equation:ME:ShannonEntropyOf-pdf}) can be referred to
        as the entropy of the probability measure 
        $P$, in the sense that the measure $P$ is induced by $p$,
        i.e.,
        \begin{equation}
        \label{Equation:ME:ProbabilityMeasureInducedByaPdf}  
          P(E) = \int_{E} p(x) \, \ud \mu(x) \enspace, \:\:\:\:\:
          \forall E \in \mathfrak{M} \enspace.
        \end{equation}
        This reference is consistent\footnote{Say 
        $p$ and 
        $r$ are two pdfs and $P$ and $R$ are corresponding
        induced measures on measurable space $(X,\mathfrak{M})$ such
        that $P$ and $R$ are identical, i.e., $\int_{E} p \,
        \ud \mu = \int_{E} r \, \ud \mu$, $\forall E \in \mathfrak{M}$. Then
        we have $p \stackrel{\mathrm{a.e}}{=} r$ and hence
        $ -\int_{X} p \ln p \, \ud \mu = -\int_{X} r \ln r \, \ud
        \mu$.} because the probability measure
        $P$ can be identified {\it a.e} by the pdf $p$.
        Further, the definition of the probability measure $P$ in
        (\ref{Equation:ME:ProbabilityMeasureInducedByaPdf}), allows one
        to write entropy functional
        (\ref{Equation:ME:ShannonEntropyOf-pdf}) 
        as 
        \begin{equation}
        \label{Equation:ME:ShannonEntropyOf-PM-inducedBy-pdf}
        S(p) = - \int_{X} \frac{\ud P}{\ud \mu} \ln \frac{\ud P}{\ud
        \mu} \, \ud \mu \enspace,
        \end{equation}
        since (\ref{Equation:ME:ProbabilityMeasureInducedByaPdf})
        implies\footnote{If a 
        nonnegative measurable function $f$ induces a measure $\nu$ on
        measurable space $(X,\mathfrak{M})$ with respect to a measure
        $\mu$, defined as $\nu(E) = \int_{E} f \, \ud \mu, \:\:\: \forall E \in
        \mathfrak{M}$ then $\nu \ll \mu$. Converse is given by
        Radon-Nikodym theorem~\cite[pp.36, Theorem
          1.40(b)]{Kantorovitz:2003:IntroductionToModernAnalysis}.} $P
        \ll \mu$, and pdf $p$ is the
        Radon-Nikodym derivative of $P$ w.r.t $\mu$. 

        Now we proceed to the definition of Kullback-Leibler
        relative-entropy or KL-entropy for probability measures.
        \begin{definition}
        \label{Definition:ME:RelativeEntropy_1}
        Let $P$ and $R$ be two probability measures on measurable
        space $(X,\mathfrak{M})$. Kullback-Leibler relative-entropy
        of $P$ relative to $R$ is defined as
        \begin{equation}
        \label{Equation:ME:RelativeEntropyOfProbabilityMeasures}
        I(P\|R) = \left\{ \begin{array}{ll}
        \displaystyle{\int_{X} \ln \frac{\ud P}{\ud R} \, \ud P }     &
        \:\:\:\:\:\textrm{if}\:\:\:\:\:  P \ll R    , \\ \\
          +\infty   & \:\:\:\:\:\textrm{otherwise.}
           \end{array} \right.
        \end{equation}
        \end{definition}

        The divergence inequality
        $I(P\|R) \geq 0$ and $I(P\|R) =0$ if and only if $P=R$ can be
        shown in this case too.
        Relative-entropy~(\ref{Equation:ME:RelativeEntropyOfProbabilityMeasures})
        also can be written as 
        \begin{equation}
        \label{Equation:ME:AnotherFormForRelativeEntropyOfProbabilityMeasures}  
        I(P\|R) = \int_{X} \frac{\ud P}{\ud R} \ln \frac{\ud P}{\ud R}
        \, \ud R \enspace.
        \end{equation}

	Let the $\sigma$-finite measure $\mu$ on $(X,\mathfrak{M})$
        such that $P \ll R \ll \mu$.
	Then
        (\ref{Equation:ME:RelativeEntropyOfProbabilityMeasures}) can
        be written as
        \begin{equation}
         \label{Equation:ME:RelativeEntropy_of_pdf} 
        I(p\|r) = \int_{X} p(x) \ln \frac{p(x)}{r(x)} \, \ud \mu(x) \enspace,
        \end{equation}
        provided the integral on right exists. The pdfs $p(x)$ and $r(x)$
        in (\ref{Equation:ME:RelativeEntropy_of_pdf})
        are the Radon-Nikodym derivatives of $P$ and $R$ with
        respect to $\mu$, i.e., $p =\frac{\ud P}{\ud \mu}$ and 
        $r=\frac{\ud R}{\ud \mu}$.
        Here in the sequel we use the convention
        \begin{equation}
        \ln 0 = - \infty, \:\:\ln \frac{a}{0} = + \infty\:
        \mathrm{for any}\: a \in \mathbb{R}, \:\:
        0.(\pm \infty) = 0.
        \end{equation}


	Shannon entropy
        in~(\ref{Equation:ME:ShannonEntropyOf-PM-inducedBy-pdf}) is
        defined for a probability measure that is induced by a pdf. 
        By the Radon-Nikodym theorem, one can
        define Shannon entropy for any arbitrary $\mu$-continuous
        probability measure as follows. 
        \begin{definition}
         \label{Definition:ME:ShannonEntropy_of_ProbabiliyMeasure} 
         Let $(X,\mathfrak{M},\mu)$ be a $\sigma$-finite measure
        space. Entropy of any $\mu$-continuous probability measure $P$
        ($P \ll \mu$) is defined as
        \begin{equation}
        \label{Equation:ME:ShannonEntropy_of_ProbabilityMeasure}  
        S(P) = - \int_{X} \ln \frac{\ud P}{\ud \mu} \, \ud P  \enspace.
        \end{equation}
        \end{definition}

        Properties of entropy of a probability measure in the
        Definition~\ref{Definition:ME:ShannonEntropy_of_ProbabiliyMeasure} are
        studied in detail by
        Ochs~\cite{Ochs:1976:BasicPropertiesOfTheGeneralizedBoltzmann-Gibbs-ShannonEntropy}.
        In the literature, one can find notation of the form 
        $S(P|\mu)$ to represent the entropy functional in
        (\ref{Equation:ME:ShannonEntropy_of_ProbabilityMeasure}) viz.,
        the entropy of a  
        probability measure, to stress the role of the measure
        $\mu$ (for
        example~\cite{Ochs:1976:BasicPropertiesOfTheGeneralizedBoltzmann-Gibbs-ShannonEntropy,Athreya:1994:EntropyMaximization}). Since
        all the information measures we define are with 
        respect to the measure $\mu$ on $(X, \mathfrak{M})$, we omit
        $\mu$ in the entropy functional notation.

        By assuming $\mu$ as a probability measure in the
        Definition~\ref{Definition:ME:ShannonEntropy_of_ProbabiliyMeasure}
        one can relate Shannon entropy with Kullback-Leibler entropy
        as
        \begin{equation}
        \label{Equation:ME:RelationBetweenMeasureTheoreticEntropyAndKullback} 
        S(P) = - I(P\|\mu).
        \end{equation}
	Note that when $\mu$ is not a probability measure, the
        divergence inequality $I(P\|\mu) \geq 0$ 
	need not be satisfied.

        Before we conclude this section, we make a note on the
        $\sigma$-finiteness of measure $\mu$. In the measure-theoretic
        definitions of Shannon entropy we assumed that $\mu$ is a
        $\sigma$-finite 
        measure. This condition was used by
        Ochs~\cite{Ochs:1976:BasicPropertiesOfTheGeneralizedBoltzmann-Gibbs-ShannonEntropy},
        Csisz\'{a}r~\cite{Csiszar:1969:OnGeneralizedEntropy} 
        and
        Rosenblatt-Roth~\cite{Rosenblatt-Roth:1964:TheConceptOfEntropyInProbabilityTheory} 
        to tailor the measure-theoretic definitions. For all practical
        purposes and for most applications this assumption is
        satisfied. (See
        \cite{Ochs:1976:BasicPropertiesOfTheGeneralizedBoltzmann-Gibbs-ShannonEntropy}
        for a discussion on the physical interpretation of measurable space
        $(X,\mathfrak{M})$ with $\sigma$-finite measure $\mu$ for
        entropic measure of the
        form~(\ref{Equation:ME:ShannonEntropy_of_ProbabilityMeasure}),
        and relaxation $\sigma$-finiteness
        condition.)
	By relaxing this condition, more universal
        definitions of entropy functionals are studied by
        Masani~\cite{Masani:1992:TheMeasureTheoreticAspectsOfEntropy_Part_1,Masani:1992:TheMeasureTheoreticAspectsOfEntropy_Part_2}.

\section{Measure-Theoretic Definitions of Generalized Information
  Measures}
\label{Section:MeasureTheoreticDefinitionsOfGeneralizedInformationMeasures}
        \noindent
	We begin with a brief 
	note on the notation and assumptions used.
	We define all the information measures 
        on the measurable space $(X,\mathfrak{M})$, and default reference
        measure is $\mu$ unless otherwise stated.
        To avoid clumsy formulations, we will not
        distinguish between functions differing on a $\mu$-null set
        only; nevertheless, we can work with equations between
        $\mathfrak{M}$-measurable functions on $X$ if they are
        stated as valid as being only $\mu$-almost everywhere ($\mu$-a.e or
        a.e).
        Further we assume that all the quantities of interest
        exist and assume, implicitly, the $\sigma$-finiteness of $\mu$ and
        $\mu$-continuity of probability measures when ever
        required. Since these assumptions repeatedly occur in various
        definitions and formulations, these will not be mentioned in
	the sequel.
        With these assumptions we do not distinguish between 
        an information measure of pdf $p$ and of corresponding probability
        measure $P$ -- hence we give definitions of
        information measures for pdfs, we use  corresponding
        definitions of probability measures as well, when ever it is
        convenient or required  --  with the understanding that $P(E) = \int_{E} p\,
        \ud \mu $, the converse being due to the Radon-Nikodym theorem, where $p =
        \frac{\ud P}{\ud \mu}$. 

	Similar to the definition of Shannon entropy
        (\ref{Equation:ME:ShannonEntropyOf-pdf}) one can extend the
        R\'{e}nyi entropy in the discrete case
        (\ref{Equation:Definition_RenyiEntropy}) to 
        measure-theoretic case as follows.
        \begin{definition}
        \label{Definition:ME:Measure-TheoreticRenyiEntropy}
        R\'{e}nyi entropy
        of a pdf $p:X \rightarrow {\mathbb{R}}^{+}$ 
        on $(X,\mathfrak{M},\mu)$ is defined as
        \begin{equation}
        \label{Equation:ME:RenyiEntropyOf-pdf}  
        S_{\alpha}(p) = \frac{1}{1-\alpha} \ln 
        \int_{X}p(x)^{\alpha}\, \ud \mu(x)  
        \enspace, 
        \end{equation}
        provided the integral on the right exists and $\alpha \in
        \mathbb{R}$ and $\alpha > 0$.
        \end{definition}

        The same can be written for any $\mu$-continuous probability
        measures $P$ as 
        \begin{equation}
        \label{Equation:ME:RenyiEntropyOf-PM}
          S_{\alpha}(P) = \frac{1}{1-\alpha} \ln \int_{X}
          {\left( \frac{\ud P}{\ud \mu} \right)}^{\alpha -1} \, \ud P
        \end{equation}  

        On the other hand, R\'{e}nyi relative-entropy can be defined as
        follows.
        \begin{definition}
        Let $p,r:X \rightarrow
        {\mathbb{R}}^{+}$ be two pdfs defined on $(X,\mathfrak{M},\mu)$. 
        R\'{e}nyi relative-entropy of $p$ relative to $r$ 
        is defined as
        \begin{equation}
        \label{Equation:ME:RenyiRelativeEntropyOf-pdf}              
        I_{\alpha}(p\|r) = \frac{1}{\alpha -1} \ln \int_{X}
        \frac{p(x)^{\alpha}}{r(x)^{\alpha -1}} \, \ud \mu(x) \enspace,
        \end{equation}
        provided integral on the right exists.
        \end{definition}
        The same can be written in terms of probability measures as
        \begin{eqnarray}
	\label{Equation:ME:RenyiRelativeEntropyOf-PMs}
          I_{\alpha}(P\|R) &=& \frac{1}{\alpha -1} \ln   \int_{X}
          {\left( \frac{\ud P}{\ud R} \right)}^{\alpha -1} \, \ud P
          \nonumber \\
          &=& \frac{1}{\alpha -1} \ln \int_{X}
          {\left( \frac{\ud P}{\ud R} \right)}^{\alpha} \, \ud R
          \enspace,
        \end{eqnarray}
	whenever $P \ll R$; $I_{\alpha}(P \|R) = + \infty$, otherwise.
       Further if we assume $\mu$ in
        (\ref{Equation:ME:RenyiEntropyOf-PM}) is a probability measure
        then 
	\begin{equation}
	\label{Equation:ME:Renyi_EntropyandRelativeEntropy}
	S_{\alpha}(P) = I_{\alpha}(P\|\mu) \enspace.
	\end{equation}

       On the other hand, it is well
       known that unlike Shannon entropy, Kullback-Leibler
        relative-entropy in the discrete 
       case can be extended naturally to the measure-theoretic case, in
       the  sense that measure-theoretic definitions can 
        be defined as a limit of a sequence of finite discrete
        entropies of pmfs which approximate the pdfs involved. This
       fact is shown for R\'{e}nyi relative-entropy in the continuous
       valued space $\mathbb{R}$ by
       R\'{e}nyi~\cite{Renyi:1960:SomeFundamentalQuestionsOfInformationTheory},
       which can be extended to the measure-theoretic case (see~\cite{DukkipatiMurtyBhatnagar:2006:OnMeasureTheoreticDefinitionsOfGeneralizedInformationMeasures}).

\section{Gelfand-Yaglom-Perez Theorem in the General Case}
\label{Section:GelfandYaglomPerez_Theorem}
        \noindent
	In the ergodic approach of information theory, basic
	definitions of information measures are given for measurable
	partitions.
	Before we proceed to the definitions we give our notation.
	Let $(X,\mathfrak{M})$ be a measurable space and $\Pi$ 
        denote the set of all measurable partitions of $X$.
	We denote a measurable partition $\pi \in \Pi$ as $\pi =
	\{E_{k}\}_{k=1}^{m}$, i.e, $\cup_{k=1}^{m} E_{k} = X$ and
	$E_{i} \cap E_{j} = \emptyset,\: i \neq j, \: i,j = 1,\ldots m$. 
	We denote the set of all simple functions on
	$(X,\mathfrak{M})$ by ${\mathbb{L}}_{0}^{+}$, and the set of all 
        nonnegative $\mathfrak{M}$-measurable functions by
        ${\mathbb{L}}^{+}$. The set of all $\mu$-integrable functions, where
        $\mu$ is a measure defined on $(X,\mathfrak{M})$, is denoted
        by $L^{1}(\mu)$. R\'{e}nyi relative-entropy $I_{\alpha}(P\|R)$
	refers to (\ref{Equation:ME:RenyiRelativeEntropyOf-PMs}), which
	can be written as
	\begin{equation}
	\label{Equation:ME:RenyiRelative-Entropy_InTermsOf_Varphi}
	I_{\alpha}(P\|R) = \frac{1}{\alpha-1} \ln \int_{X}
	\varphi^{\alpha}\, \ud R \enspace,
	\end{equation}
	where $\varphi \in L^{1}(R)$ is defined as $\varphi =
	\frac{\ud P}{\ud R}$.

        Let $P$ and $R$ be two
        probability measures on $(X, \mathfrak{M})$ such that $P \ll
        R$. Relative entropy of partition $\pi \in \Pi$ with $P$ with
        respect to $R$ is defined as
        \begin{equation}
          I_{P\|R}(\pi) = \sum_{k=1}^{m} P(E_{k}) \ln
        \frac{P(E_{k})}{R(E_{k})}\enspace. 
         \end{equation}
	Now, the GYP-theorem for KL-entropy states that
	\begin{equation}
	 I(P\|R) = \sup_{\pi \in \Pi}  I_{P\|R}(\pi) \enspace,
	\end{equation}
	where $I(P\|R)$ measure-theoretic KL-entropy defined as in
        Definition~\ref{Definition:ME:RelativeEntropy_1}. When $P$ is
        not absolutely continuous with respect to $R$, GYP-theorem
        assigns $I(P\|R) = + \infty$.  The proof of
        GYP-theorem given by
        Dobrushin~\cite{Dobrushin:1959:GeneralFormulationsOfShannonsbasicTheorems}
        can be found in~\cite[pp. 23, Theorem
          2.4.2]{Pinsker:1960:InformationAndInformationStability} or
        in~\cite[pp. 92, Lemma 5.2.3]{Gray:1990:EntropyAndInformationTheory}.
  \subsection{GYP for R\'{e}nyi Relative-Entropy}
	\noindent
	Before we state and prove the GYP-theorem for R\'{e}nyi
	relative-entropy of order $\alpha >1$, we state the following lemma.
        \begin{lemma}
         \label{Lemma:ME:LemmaForMeasurablePartionTheorm} 
          Let $P$ and $R$ be probability measures on the measurable
          space $(X,\mathfrak{M})$ such that $P \ll R$. Let $\varphi =
          \frac{\ud P}{\ud R}$. Then for any
          $E \in \mathfrak{M}$ and $\alpha > 1$ we have
          \begin{equation}
           \label{Equation:ME:ProofUsingHoldersInequality} 
            \frac{P(E)^{\alpha}}{R(E)^{\alpha -1}} \leq \int_{E}
            \varphi^{\alpha} \, \ud R \enspace.
          \end{equation}
         \end{lemma}
         \proof
         Since $P(E) = \int_{E} \varphi \, \ud R$, $\forall E
	\in \mathfrak{M}$,
         by H\"{o}lder's inequality we have
         \begin{displaymath}
         \int_{E} \varphi  \, \ud R \leq  {\left( \int_{E} \varphi^{\alpha}\,
             \ud R \right)}^{\frac{1}{\alpha}} {\left( \int_{E} \, \ud R\right)}^{1 -
         \frac{1}{\alpha}} \enspace.  
         \end{displaymath}
         That is 
         \begin{displaymath}
           P(E)^{\alpha} \leq {R(E)}^{\alpha(1-\frac{1}{\alpha})}   \int_{E} \varphi^{\alpha}\,
             \ud R \enspace,
         \end{displaymath}
         and hence (\ref{Equation:ME:ProofUsingHoldersInequality}) follows.
         Since $P \ll R$, it is clear that this inequality reduces
            to $0 = 0$ if $R(E) = 0$.
         \endproof

         First we present our main result in its special case as follows.
         \begin{lemma}
          \label{Lemma:GYP-TheremInGeneralCase_ForSimpleFunction}  
	   Let $P$ and
          $R$ be two probability measures such that $P \ll R$. Let
          $\varphi = \frac{\ud P}{\ud R} \in {\mathbb{L}}_{0}^{+}$. Then
          for any $ 0 < \alpha < \infty $, we have
          \begin{equation}
	  \label{Equation:InLemma_GYP-TheremInGeneralCase_ForSimpleFunction_Statement}
          I_{\alpha}(P \| R) = 
          \frac{1}{\alpha-1} \ln \sum_{k=1}^{m}
         \frac{{P(E_{k})}^{\alpha}}{{R(E_{k})}^{\alpha 
          -1}} \enspace,
          \end{equation}
          where  $ \{E_{k}\}_{k=1}^{m}\in \Pi$ is the measurable
          partition corresponding to $\varphi$.
         \end{lemma}
         \proof
         The simple function 
         $\varphi \in {\mathbb{L}}_{0}^{+}$ can be written as
         $\varphi(x) = \sum_{k=1}^{m} a_{k} \chi_{E_{k}}(x)$, $\forall
         x \in X $, 
         where $a_{k} \in \mathbb{R}$, $k=1, \ldots m$.
         Now we have
         $P(E_{k}) = \int_{E_{k}} \varphi \, \ud R = a_{k} R(E_{k})$,
         and hence
         \begin{equation}
	 \label{Equation:ME:InLemma_GYP-TheremInGeneralCase_ForSimpleFunction_Int1}
         a_{k} = \frac{P(E_{k})}{R(E_{k})} \enspace,\:\:\:\:\:\:
         \forall k =1, \ldots m . 
         \end{equation}
         We also have
         $\varphi^{\alpha}(x) = \sum_{k=1}^{m} a_{k}^{\alpha}
         \chi_{E_{k}}$, $\forall x \in X $
         and hence
         \begin{equation}
	 \label{Equation:ME:InLemma_GYP-TheremInGeneralCase_ForSimpleFunction_Int2}
           \int_{X} \varphi^{\alpha}\, \ud R = \sum_{k=1}^{m}
           a_{k}^{\alpha} R(E_{k}) \enspace.
         \end{equation}
	Now, from
         (\ref{Equation:ME:RenyiRelative-Entropy_InTermsOf_Varphi}),
         (\ref{Equation:ME:InLemma_GYP-TheremInGeneralCase_ForSimpleFunction_Int1})
         and
         (\ref{Equation:ME:InLemma_GYP-TheremInGeneralCase_ForSimpleFunction_Int2})
         one obtains
         (\ref{Equation:InLemma_GYP-TheremInGeneralCase_ForSimpleFunction_Statement}).  
         \endproof
	Now we state and prove GYP-theorem for R\'{e}nyi relative-entropy.
         \begin{theorem}
         \label{Theorm:GYP-TheremInGeneralCase}  
          Let $(X,\mathfrak{M})$ be a measurable space and $\Pi$
          denote the set of all measurable partitions of $X$. Let $P$ and
          $R$ be two probability measures. Then
          for any $\alpha >1$, we have
          \begin{equation}
          I_{\alpha}(P \| R) = \sup_{\{E_{k}\}_{k=1}^{m} \in \Pi}
          \frac{1}{\alpha-1} \ln \sum_{k=1}^{m}
         \frac{{P(E_{k})}^{\alpha}}{{R(E_{k})}^{\alpha 
          -1}} \enspace,
          \end{equation}
	  if $P \| R$, otherwise $I_{\alpha}(P \| R) = + \infty $.
          \end{theorem}
          \proof
	   If $P$ is not absolutely continuous with respect $R$, Then
         there exists $E \in \mathfrak{M}$ such that $P(E) > 0$ and
         $R(E) =0$. Since $\{E, X -E \} \in \Pi$, $I_{\alpha}(P \| R)
         = + \infty$.

	 Now, we assume that $P \ll R$.
            It is clear that it is enough to prove that
            \begin{equation}
             \label{Equation:ME:GenGYP-Thm_ItIsEnoughProve_1} 
              \int_{X} \varphi^{\alpha} \, \ud R = \sup_{\{E_{k}\}_{k=1}^{m} \in \Pi}\:\:
               \sum_{k=1}^{m}
               \frac{{P(E_{k})}^{\alpha}}{{R(E_{k})}^{\alpha-1}} \enspace, 
            \end{equation}  
            where $\varphi = \frac{\ud P}{\ud R}$.
            From Lemma~\ref{Lemma:ME:LemmaForMeasurablePartionTheorm},
            for any measurable partition $\{E_{k}\}_{k=1}^{m} \in \Pi$,
            we have
            \begin{displaymath}
             \sum_{k=1}^{m}
             \frac{{P(E_{k})}^{\alpha}}{{R(E_{k})}^{\alpha-1}}
             \leq
             \sum_{k=1}^{m} \int_{E_{k}} \varphi^{\alpha} \, \ud R =
             \int_{X} \varphi^{\alpha} \, \ud R \enspace,
            \end{displaymath}  
            and hence
            \begin{equation}
             \sup_{\{E_{k}\}_{k=1}^{m} \in \Pi}\:\:
               \sum_{k=1}^{m}
               \frac{{P(E_{k})}^{\alpha}}{{R(E_{k})}^{\alpha-1}}
               \leq \int_{X} \varphi^{\alpha} \, \ud R \enspace.
            \end{equation}
            Now we shall obtain the  reverse inequality to prove
            (\ref{Equation:ME:GenGYP-Thm_ItIsEnoughProve_1}) .
            That is we shall obtain 
            \begin{equation}
            \label{Equation:ME:GenGYP-Thm_ItIsEnoughProve_2}  
             \sup_{\{E_{k}\}_{k=1}^{m} \in \Pi}\:\:
               \sum_{k=1}^{m}
               \frac{{P(E_{k})}^{\alpha}}{{R(E_{k})}^{\alpha-1}}
               \geq \int_{X} \varphi^{\alpha} \, \ud R \enspace.
            \end{equation}

            Note that corresponding to any $\varphi \in {\mathbb{L}}^{+}$,
            there exists a sequence of simple functions
            $\{\varphi_{n}\}$, $\varphi_{n} \in {\mathbb{L}}_{0}^{+}$,
            which satisfies
            \begin{equation}
              0 \leq \varphi_{1} \leq \varphi_{2} \leq \ldots \leq \varphi
            \end{equation}
            such that $\lim_{n \to \infty} \varphi_{n} = \varphi$ (see
            \cite[Theorem
            1.8(2)]{Kantorovitz:2003:IntroductionToModernAnalysis}).
            $\{\varphi_{n}\}$ induces a sequence of measures 
            $\{P_{n}\}$ on $(X,\mathfrak{M})$ defined by
            \begin{equation}
              P_{n}(E) = \int_{E} \varphi_{n}(x) \, \ud R(x)\enspace,
              \:\:\:\:\: \forall E \in \mathfrak{M}.
            \end{equation}  
            We have $\int_{E} \varphi_{n} \, \ud R \leq \int_{E}
            \varphi \, \ud R < \infty, \forall E \in \mathfrak{M}$ and hence $P_{n} \ll R,
            \:\:\forall n$. From the Lebesgue bounded convergence
            theorem, we have
            \begin{equation}
              \lim_{n \to \infty} P_{n}(E) = P(E) \enspace, \:\:\:\:\: \forall E
              \in \mathfrak{M} \enspace.
            \end{equation}
            Now, $\varphi_{n} \in \mathbb{L}_{0}^{+} $, $\varphi_{n}^{\alpha} \leq \varphi_{n+1}^{\alpha} \leq
            \varphi^{\alpha}$, $1\leq n < \infty$ and $\lim_{n \to
              \infty} \varphi_{n}^{\alpha} = \varphi^{\alpha}$  for
            any $\alpha > 0$. Hence
            from Lebesgue monotone convergence theorem
            \cite[pp.21]{Rudin:1966:RealAndComplexAnalysis} we have 
            \begin{equation}
             \label{Equation:ME:GenGYP-Thm_Intermediate_1} 
              \lim_{n \to \infty} \int_{X}\varphi_{n}^{\alpha} \, \ud R
              = \int_{X} \varphi^{\alpha} \, \ud R \enspace.
            \end{equation}  
            The claim is that 
            (\ref{Equation:ME:GenGYP-Thm_Intermediate_1}) implies
            \begin{equation}
            \label{Equation:ME:GenGYP-Thm_Intermediate_2}               
              \int \varphi^{\alpha}\, \ud R =
              \sup \left\{ \int_{X} \phi \, \ud R \, | \, 0 \leq \phi \leq
              \varphi^{\alpha}\, , \phi \in {\mathbb{L}}_{0}^{+}
            \right\} \enspace.
            \end{equation}
            This can be verified as follows.
            Denote $\phi_{n} = \varphi_{n}^{\alpha}$. We have $0 \leq
            \phi \leq \varphi^{\alpha}$, $\forall n$, $\phi_{n}
            \uparrow \varphi^{\alpha}$, and
            \begin{equation}
            \label{Equation:ME:GenGYP-Thm_Intermediate_2_1}               
              \lim_{n \to \infty}  \int_{X} \phi_{n}\, \ud R =
              \int_{X} \varphi^{\alpha} \, \ud R \enspace.
            \end{equation}  
           For any $\phi \in {\mathbb{L}}_{0}^{+} $ such that $0 \leq
            \phi \leq \varphi^{\alpha}$ we have
            \begin{displaymath}
              \int_{X} \phi \, \ud R \leq \int_{X} \varphi^{\alpha} \,
              \ud R
            \end{displaymath}  
           and hence
           \begin{equation}
            \label{Equation:ME:GenGYP-Thm_Intermediate_5}
              \sup \left\{ \int_{X} \phi \, \ud R \, | \, 0 \leq \phi \leq
              \varphi^{\alpha}\, , \phi \in {\mathbb{L}}_{0}^{+}
              \right\} \leq
              \int \varphi^{\alpha}\, \ud R \enspace.
           \end{equation}
           Now we get reverse inequality of
           (\ref{Equation:ME:GenGYP-Thm_Intermediate_5}). 
           If $\int_{X} \varphi^{\alpha} \, \ud R < +\infty$, from
           (\ref{Equation:ME:GenGYP-Thm_Intermediate_2_1}) 
           given any $\epsilon > 0$ one can find $0 \leq n_{0} <
           \infty$ such that
           \begin{displaymath}
             \int_{X} \varphi^{\alpha} \, \ud R < \int_{X} \phi_{n_{0}} \, \ud
             R + \epsilon
           \end{displaymath}
           and hence
           \begin{equation}
           \label{Equation:ME:GenGYP-Thm_Intermediate_6}
             \int_{X} \varphi^{\alpha} \, \ud R < \sup \left\{
             \int_{X} \phi \, \ud R \, | \, 0 \leq \phi \leq 
              \varphi^{\alpha}\, , \phi \in {\mathbb{L}}_{0}^{+}
              \right\} + \epsilon \enspace.
           \end{equation}
           Since (\ref{Equation:ME:GenGYP-Thm_Intermediate_6})
           is true for any $\epsilon >  0$ we can write
           \begin{equation}
           \label{Equation:ME:GenGYP-Thm_Intermediate_7}
             \int_{X} \varphi^{\alpha} \, \ud R \leq \sup \left\{
             \int_{X} \phi \, \ud R \, | \, 0 \leq \phi \leq 
              \varphi^{\alpha}\, , \phi \in {\mathbb{L}}_{0}^{+}
              \right\} \enspace.
           \end{equation}
           Now let us verify
           (\ref{Equation:ME:GenGYP-Thm_Intermediate_7}) in the
           case of 
           $\int_{X} \varphi^{\alpha} \, \ud R = +\infty$. In this
           case,  
           $\forall N > 0$, one can choose $n_{0}$ such that
           $\int_{X} \phi_{n_{0}} \, \ud R > N$ and hence
           \begin{equation}
           \label{Equation:ME:GenGYP-Thm_Intermediate_8_1}  
             \int_{X} \varphi^{\alpha} \, \ud R > N \:\:\:\:\:\:\:\:\:\:\:\:\: (\because 0
             \leq \phi_{n_{0}} \leq \varphi^{\alpha})
           \end{equation}
           and
           \begin{equation}
           \label{Equation:ME:GenGYP-Thm_Intermediate_8_2}
             \sup \left\{
             \int_{X} \phi \, \ud R \, | \, 0 \leq \phi \leq 
              \varphi^{\alpha}\, , \phi \in {\mathbb{L}}_{0}^{+}
              \right\} > N \enspace.
           \end{equation}  
           Since (\ref{Equation:ME:GenGYP-Thm_Intermediate_8_1})
           and (\ref{Equation:ME:GenGYP-Thm_Intermediate_8_2})
           are true for any $N > 0$ we have
           \begin{equation}
           \label{Equation:ME:GenGYP-Thm_Intermediate_9}
           \int_{X} \varphi^{\alpha} \, \ud R =
           \sup \left\{
             \int_{X} \phi \, \ud R \, | \, 0 \leq \phi \leq 
              \varphi^{\alpha}\, , \phi \in {\mathbb{L}}_{0}^{+}
              \right\} = + \infty
            \end{equation} 
            and hence
            (\ref{Equation:ME:GenGYP-Thm_Intermediate_7}) is
            verified in the case of  $\int_{X} \varphi^{\alpha} \, \ud
            R = +\infty$. 
           Now (\ref{Equation:ME:GenGYP-Thm_Intermediate_5})
           and (\ref{Equation:ME:GenGYP-Thm_Intermediate_7})
           verifies the claim that
           (\ref{Equation:ME:GenGYP-Thm_Intermediate_1}) implies
           (\ref{Equation:ME:GenGYP-Thm_Intermediate_2}).
           Finally (\ref{Equation:ME:GenGYP-Thm_Intermediate_2})
           together with the
           Lemma~\ref{Lemma:GYP-TheremInGeneralCase_ForSimpleFunction}
           proves
           (\ref{Equation:ME:GenGYP-Thm_ItIsEnoughProve_1}) and
           hence the theorem.
          \endproof
  \subsection{GYP for Tsallis Relative-Entropy}
	\noindent
	Due to an increasing interest in long-range correlated systems
	and non-equilibrium phenomena there has recently been much
	focus on the Tsallis (or nonextensive)
	entropy. Although, first introduced by Havrda and Charv{\'{a}}t
	\cite{HavrdaCharvat:1967:QuantificationMethodOfClassificationProcess}
	in the context of cybernetics theory 
        and later studied by
	Dar{\'{o}}czy~\cite{Daroczy:1970:GeneralizedInformationFunctions},
	it was 
	Tsallis~\cite{Tsallis:1988:GeneralizationOfBoltzmannGibbsStatistics}
	who exploited its nonextensive features and placed it in a
	physical setting. Tsallis entropy of a pdf $p$ defined on
            $(X,\mathfrak{M},\mu)$ can be defined as,
	\begin{equation}
        \label{Equation:ME:TsallisEntropyOf-pdf}  
        S_{q}(p) = \int_{X} p(x) \ln_{q} \frac{1}{p(x)}\, \ud \mu(x) =
        \frac{1 - \int_{X} p(x)^{q}\, \ud \mu(x) }{q-1}
        \enspace, 
        \end{equation}
        provided the integral on the right exists and $q \in
        \mathbb{R}$, and $q > 0$. $\ln_{q}$ in
            (\ref{Equation:ME:TsallisEntropyOf-pdf}) is referred to as
            $q$-logarithm and is defined as $\ln_{q} x = \frac{\displaystyle
            x^{1-q} -1}{\displaystyle 1-q} 
        \:\:\: (x >0, q \in {\mathbb{R}})$.
	Tsallis entropy too, like R\'{e}nyi entropy, is a
	one-parameter generalization of 
	Shannon entropy in the sense that $q \rightarrow 1$ in
	(\ref{Equation:ME:TsallisEntropyOf-pdf}) retrieves Shannon
	entropy. Tsallis entropy can be defined for$\mu$-continuous probability
        measure $P$ can be written as 
        \begin{equation}
	\label{Equation:ME:TsallisEntropyOf-PM}
           S_{q}(P) = \int_{X} \ln_{q}  {\left(\frac{\ud P}{\ud \mu}\right)}^{-1}
          \, \ud P \enspace.
        \end{equation} 

	In this framework, Tsallis relative-entropy is defined as
	\begin{equation}
        \label{Equation:ME:TsallisRelativeEntropyOf-pdf}            
        I_{q}(p\|r) = - \int_{X} p(x) \ln_{q} \frac{r(x)}{p(x)}\, \ud
        \mu(x)    = \frac{\int_{X} \frac{p(x)^{q}}{r(x)^{q-1}}\,
          \ud \mu -1 }{q-1} \enspace,
        \end{equation}
        provided all the integrals mentioned above exist and $q \in
        \mathbb{R}$, and $q > 0$.
	The same can be written for two probability measures $P$ and
        $R$ as
        \begin{equation}
        \label{Equation:ME:TsallisRelativeEntropyOf-PMs}            
          I_{q}(P\|R)= - \int_{X} \ln_{q} {\left(\frac{\ud P}{\ud R}\right)}^{-1}\,
          \ud P \enspace,
        \end{equation}
	whenever $P \ll R$; $I_{q}(P \|R) = + \infty$, otherwise.
	If $\mu$ in
        (\ref{Equation:ME:TsallisEntropyOf-PM}) is a probability measure
        then we have
	\begin{equation}
	\label{Equation:ME:Tsallis_EntropyandRelativeEntropy}
	S_{q}(P) = I_{q}(P\|\mu) \enspace.
	\end{equation}

	Now, from the fact that R\'{e}nyi and Tsallis relative-entropies
	((\ref{Equation:ME:RenyiRelativeEntropyOf-PMs}) and
	(\ref{Equation:ME:TsallisRelativeEntropyOf-PMs}) respectively)
	are monotone and continuous functions of each other, the
	GYP-theorem presented in the case of R\'{e}nyi is valid for the
	Tsallis case too, whenever $q >1$.

\section{Conclusions}
\label{Section:Conclusions}
	\noindent
	Relative-entropy or KL-entropy is an important concept in
	information theory, 
	since information measures like entropy and
	mutual information can be formulated as special
	cases. Further, KL-entropy overcomes the shortcomings of
	entropy in non-discrete settings. Note that all the above hold
	even for generalized information measures.

	GYP-theorem provides a means to
	compute KL-entropy and studying its
	behavior~\cite{Gray:1990:EntropyAndInformationTheory}. In this 
	paper, we presented the measure-theoretic definitions of
	generalized information measures. We stated and proved the
	GYP-theorem for generalized relative entropies
	of order $\alpha >1$ ($q >1$ for the  Tsallis
	case). However, results are yet to be achieved for the case $0
	< \alpha < 1$.


\section*{References}

\bibliographystyle{unsrt}
\bibliography{papi}

\end{document}